# Investigation of the thermal stability of Mg/Co periodic multilayers for EUV applications

M.-H. Hu[1], K. Le Guen[1], J.-M. André[1], S. K. Zhou[2], H. Ch. Li[2], J. T. Zhu[2], Z. S. Wang[2], C. Meny[3], N. Mahne[4], A. Giglia[4], S. Nannarone[4,5], I. Estève[6], M. Walls[7], P. Jonnard[1]

*(1)* Laboratoire Chimie Physique – Matière Rayonnement, UPMC Univ Paris 06, CNRS UMR 7614, 11 rue Pierre et Marie Curie, F-75231 Paris cedex 05, France

*(2)* Institute of Precision Optical Engineering, Department of Physics, Tongji University, Shanghai 200092, P.R. China

*(3)* Institut de Physique et Chimie des Matériaux de Strasbourg, CNRS UMR 7504, 23 rue du Loess, BP 43, F-67034 Strasbourg cedex 2, France

*(4)* CNR-IOM Laboratorio TASC, SS 14 km 163,5, I-34149 Basovizza, Trieste, Italy

(5) Università di Modenae R.E., Dipartimento Ingegneria Materiali, viaVignolese 905/a, Modena, Italy

*(6)*Institut de Minéralogie et de Physique des Milieux Condensés, Univ Paris 06 et 07, CNRS UMR 7590, 4 place Jussieu, F-75252 Paris Cedex 05, France

*(7)* Laboratoire de Physiques des Solides, CNRS UMR 8502, Univ Paris Sud, F-91405 Orsay, France

---

[1] Corresponding author: Dr. Philippe Jonnard, Laboratoire Chimie Physique – Matière et Rayonnement, tel: 33 1 44 27 63 03; fax: 33 1 44 27 62 26

ABSTRACT

We present the results of the characterization of Mg/Co periodic multilayers and their thermal stability for the EUV range. The annealing study is performed up to a temperature of 400°C. Images obtained by scanning transmission electron microscopy and electron energy loss spectroscopy clearly show the good quality of the multilayer structure. The measurements of the EUV reflectivity around 25 nm (~49 eV) indicate that the reflectivity decreases when the annealing temperature increases above 300°C. X-ray emission spectroscopy is performed to determine the chemical state of the Mg atoms within the Mg/Co multilayer. Nuclear magnetic resonance used to determine the chemical state of the Co atoms and scanning electron microscopy images of cross sections of the Mg/Co multilayers reveal changes in the morphology of the stack from an annealing temperature of 305°C. This explains the observed reflectivity loss.

## 1. Introduction

In the extreme ultraviolet (EUV) range, highly efficient periodic multilayers are widely used in many fields such as X-ray space telescopes, EUV lithography or X-ray laser facilities. The previous measurements of the optical and interface properties of the Mg/Co multilayer [1, 2] show that this multilayer is promising for future applications in the wavelength range close to the Mg L edge around 25 nm (or a photon energy of 50 eV) and in particular for the observation of the 30.4 nm He II emission [3]. In the 25 nm range, a reflectivity of about 43% is measured at 45° for *s*-polarized radiation.

Another important property often required for multilayer devices is their thermal stability. In fact multilayers can be subject to high thermal loads, for example, when in use in space telescopes where they needs to withstand the high temperature space environment. Systems already investigated in this context include Mg/SiC [4,5] which is thermally stable for applications only up to 200°C and also on more stable systems such as Mo/Si, Mo/C, W/Si and Fe/Al, etc. [6-11] and more recently Mo/$B_4$C [12].

The present work examines Mg/Co multilayers and their thermal stability by studying interdiffusion, interfacial compound formation and roughness development. From the phase diagram, the Mg-Co system is stable in the solid state up to a high temperature. The only possible compound is $Co_2Mg$. In this work, we study Mg/Co multilayer samples prepared by magnetron sputtering and then annealed up to 400°C, under ultra-high vacuum. Scanning transmission electron microscopy (STEM) images and electron energy loss spectroscopy (EELS) give direct evidence of the quality of the layers of the unannealed multilayer. The measurements of the EUV reflectivity around 25 nm (~49 eV) indicate the optical property of the multilayers. X-ray emission spectroscopy (XES) and nuclear magnetic resonance (NMR) are performed to determine the chemical state of the Mg and Co atoms within the Mg/Co multilayer, respectively. The interest of combining XES and NMR is to obtain the information from both sides of the interfaces. Scanning electron microscopy (SEM) checks the changes of the interface morphology as a function of the annealing temperature.

## 2. Experimental Section

### *2.1 Samples*

Following the multilayer design resulting from simulations [13], we prepared two sets of Mg/Co multilayers with different periods, Mg/Co_1 and Mg/Co_2 whose parameters are given in Table 1. They were both deposited onto ultra-smooth polished Si substrates (30 mm × 40 mm) by an ultrahigh vacuum direct current magnetron sputtering deposition system (Model JGP560C, SKY Technology, China) using Co (99.95% purity) and Mg (99.98% purity) targets. The sample preparation was done at a base pressure of $1\times10^{-4}$ Pa. The working gas is argon (99.999% purity) at a constant working pressure 0.13 Pa. The number of periods is 30. In order to prevent oxidation, both multilayers are capped with a 3.5 nm thick $B_4C$ layer. For Mg/Co_2, we cut the 15 mm × 15 mm sample into six small pieces of size 7.5 mm × 5 mm in order to do the annealing at different temperatures.

**Table 1.** Designed structures of the Mg/Co multilayers.

| Sample name | Period d (nm) | $d_{Co}$ (nm) | $d_{Mg}$ (nm) | Simulated reflectivity | Size (mm$^2$) | Experimental | Annealing |
|---|---|---|---|---|---|---|---|
| Mg/Co_1 | 8.00 | 2.55 | 5.45 | - | 10×10 | XES | Yes |
|  |  |  |  |  | 4×6 | NMR | Yes |
| Mg/Co_2 | 17.00 | 2.55 | 14.45 | 56.7% @ 25.2 nm | 15×15 | STEM / EELS | No |
|  |  |  |  |  | 7.5×5 | EUV reflectivity | Yes |
|  |  |  |  |  | 7.5×5 | SEM | Yes |

We characterized the Mg/Co_2 unannealed multilayer after its preparation by measuring its the reflectivity at 0.154 nm of [1]. The structural parameters of the Mg/Co_2 multilayer, deduced from the fit of the reflectivity curve, show a good agreement between the aimed and effective thicknesses of the layers and an interfacial roughness of Co-on-Mg and Mg-on-Co layers which are 0.5 and 0.6 nm, respectively.

To investigate the thermal stability of Mg/Co periodic multilayers, all the Mg/Co multilayers were annealed at 280, 305 and 400°C for one hour centre in the preparation chamber of the BEAR beamline [14] at the Elettra synchrotron. The base pressure is $1 \times 10^{-8}$ Pa. All samples were annealed for one hour. The EUV reflectivity measurements were made *in situ*, that is to say that the sample remained in a UHV environment. The XES, NMR and SEM measurements were done *ex situ* after transferring the samples in air.

The Mg/Co_2 unannealed sample is dedicated to the STEM / EELS (Table 1) analysis. The Mg/Co_2 annealing samples are dedicated to the reflectivity measurement in the EUV range and SEM studies in order to characterize the optical properties and the morphology of the multilayers, respectively. Then the set of Mg/Co_1 annealing samples, which is characterized by a short period, is dedicated to the XES and NMR analysis in order to study the chemical state of the Mg and Co atoms, respectively. Indeed, to be able to identify the formation of an interfacial layer at the interface between two successive layers, the thickness of this interfacial layer should be comparable to that of the layer of the emitting atoms.

*2.2 Scanning transmission electron microscope and electron energy loss spectroscopy*

The lamella of the cross section of the unannealed Mg/Co_2 multilayer was examined by STEM and EELS. A Nion Ultrastem 100 operating at 100 kV and under ultra-high vacuum was used for these observations. The spherical aberration-corrected probe size was around 0.1 nm. The images are generated by scanning the focused beam over the lamella. Both energy-loss spectra and high angle annular dark field (HAADF) images can be collected simultaneously, which makes the STEM a good technique for materials morphological and chemical characterization at the sub-nanometer level. We used a probe convergence angle of 35 mrad and a collection angle for the EELS spectrometer of 50 mrad. Spectra were acquired with a dispersion of 1 eV per channel (with a 1024 channel camera) to have access simultaneously to the Co-L edge (779 eV) and the Mg K edge (1305 eV). The HAADF collection range was 75 to 200 mrad.

*2.3 EUV reflectivity*

The EUV reflectivity curves were measured on the BEAR beamline of the Elettra synchrotron facility using *s*-polarized light. The absolute uncertainty of the reflectivity values was about 1%. The photon energy was calibrated with respect to the Pt $4f_{7/2}$ and Si L with an accuracy of 0.1 eV. The intensities of incident and reflected radiation were measured with a solid-state photodiode. The incoming photon flux was also monitored using an Au mesh inserted into the beam path, whose drain current was used for normalization.

*2.4 Nuclear magnetic resonance spectroscopy*

In order to probe the chemical state of the Co atoms within the multilayers, the samples have been analyzed by zero field NMR. The experiment was performed at 4.2 K using an automated frequency scanning broadband spectrometer with phase coherent detection [15]. The NMR spectra represent the distribution of Co atoms versus their resonance frequency [16]. The NMR frequency is strongly dependent on the number and nature of atoms in its neighborhood and also possibly on the symmetry of this neighborhood.

*2.5 X-ray emission spectroscopy*

We have performed XES with a high-resolution wavelength dispersive spectrometer [17] to identify the chemical state of the Mg atoms within the multilayers. Similar studies have already been performed by this method [18-26]. We observed the shape of the Mg Kβ emission band from the 3p–1s transition which describes the Mg 3p valence states. They are sensitive to the chemical state of the Mg atoms [27,28]. The energy of the incident electrons used to produce the ionizations necessary to induce the electron transitions was set to 7.5 keV. The electron beam comes from a Pierce gun. Following the ionization of the atoms present in the sample, characteristic x-rays were emitted [29,30], then dispersed by a (100) bent beryl crystal. The radiation was detected in a gas-flux counter working in the Geiger regime. The spectra of the multilayers were compared to the references, Mg metal and MgO.

*2.6 Scanning electron microscopy and focused ion beam*

Lamellae of the cross section of the $Mg/Co_2$ multilayers were prepared by FIB (Focused Ion Beam) / SEM CrossBeam® workstation NEON40EsB (Carl Zeiss). The FIB/SEM system, combination of a scanning electron microscope and a $Ga^+$ ion beam, allowed a precise and located milling by sputtering of material with a simultaneous accurate observation. This FIB/SEM workstation is also equipped with an *in situ* gas injection system loaded with platinum gas precursor.

We first deposited a thin carbon rich platinum film (about 15 nm thick) with the electron beam to protect the sample surface against ion-beam damage. Then a 1 $\mu$m thick platinum layer was deposited with the ion beam. A first cross section of the multilayers is cut and observed by SEM. A lamella, whose size was approximately 10 $\mu$m × 4 $\mu$m with a thickness of 1 $\mu$m, was excavated, lifted-out *in situ* using a micromanipulator and glued to a copper half TEM grid. Finally, the lamella was polished to a thickness of 60 nm, giving the electron transparency needed for TEM. However, further thinning to about 30 nm in a Gatan PIPS system operating at 1 kV was found to be necessary for the high resolution imaging and EELS experiments.

## 3. Results and discussion

*3.1 Samples without annealing*

Figure 1 shows us the overall view of STEM images of the unannealed $Mg/Co_2$ multilayer as prepared by FIB. We can clearly distinguish the multilayer structures: from top to bottom are the substrate, the 30 periods multilayer and the Pt protection layer. The figure shows us the multilayer of good quality and the layers are smooth. The diffuse nature of the interfaces is probably mostly due to loss of resolution due to multiple scattering in the thicker parts of the TEM sample (which happens here to be mainly in the lower part of the figure, in the layers nearest the substrate).

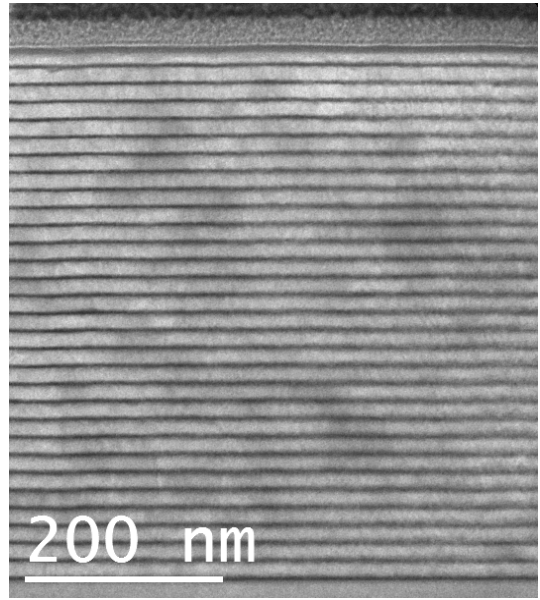

**Figure 1.** STEM BF image of the unannealed Mg/Co_2 multilayer.

To get a better impression of the quality of the layers we need to zoom onto a part thin enough (and well oriented enough) to allow visualisation of the crystal structure in the dense Co layers. Figure 2 shows a zoom onto the first two periods of the multilayer in the thinnest part of the sample. The horizontal fringes have spacings corresponding to those of the hexagonal planes in bulk magnesium and bulk cobalt in the appropriate regions. Lattice spacings are measured as $0.26 \pm 0.01$nm in the Mg layer ($c/2 = 0.2605$nm in bulk Mg) and $0.21$nm $\pm 0.01$nm in the Co layer ($c/2 = 0.204$nm in bulk Co). Other orientations are sometimes visible, but are probably due, in this thin region, to disordered and/or oxide layers at the outer surfaces of the STEM sample caused by to ion-beam damage and air exposure. (In most thicker regions, the hexagonal Mg fringes can be seen but not those of Co, which is too dense). The layers thus appear to be generally well-crystallised and somewhat epitaxial in nature. A poorly crystallised region is also seen between the substrate and the first Co layer. In order to determine the nature of this layer and to attempt a further check of the roughness of the layers some preliminary EELS

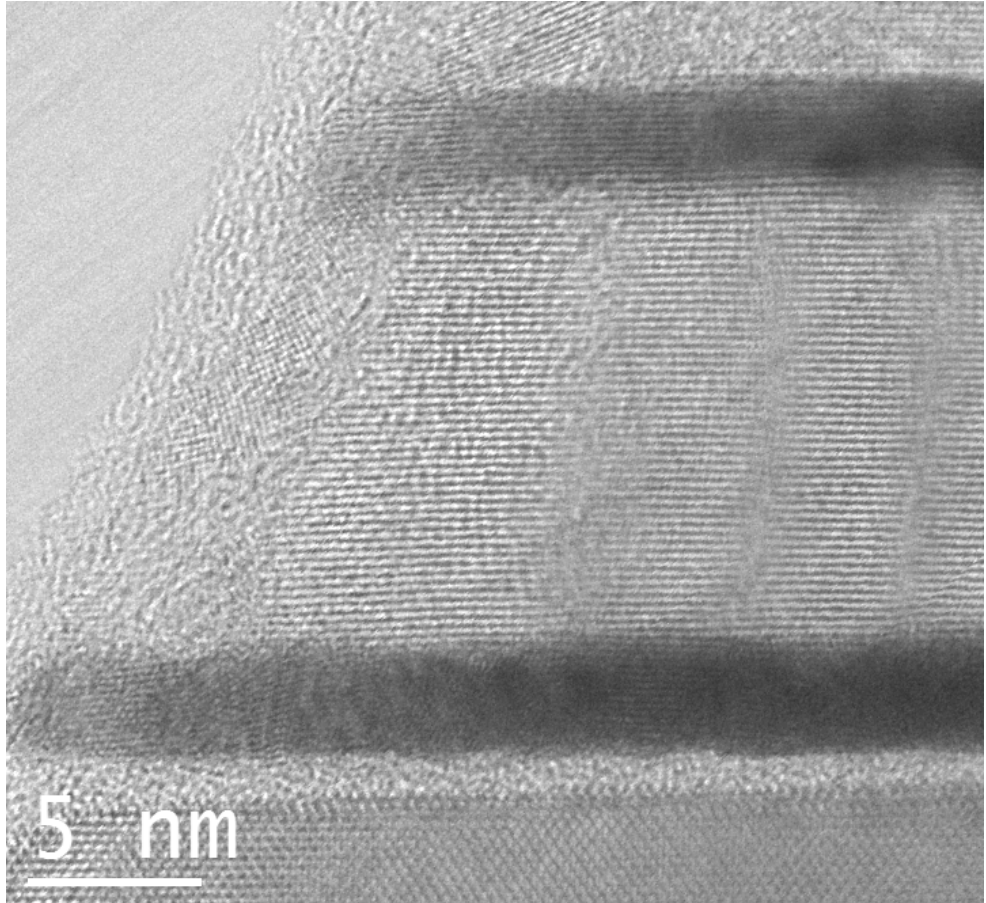

spectrum-images have been acquired in this thinner region.

**Figure 2.** Close up of the thinnest part of the unannealed Mg/Co_2 multilayer.

Figure 3 shows Co, Mg, O EELS maps and the corresponding HAADF images from the first two bi-layers. The layers can be seen to be chemically well-ordered at this scale. A strong oxygen signal is detected, along with Mg in the poorly crystallised region mentioned above. This appears therefore to be an MgO layer. The oxygen signal is also slightly enhanced over the Mg layers, probably due to surface oxidation.

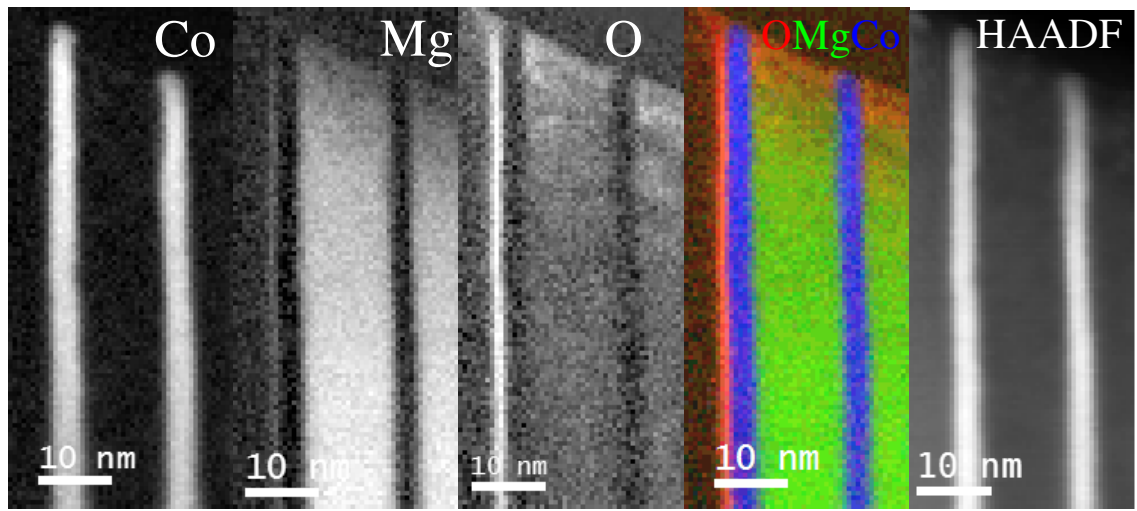

**Figure 3.** Co, Mg, O EELS maps and the corresponding HAADF images from the first two bi-layers.

Figure 4 shows a higher resolution set of EELS maps going across one Co layer. The vertical integrations in the dotted boxes give the interface width. The width of the interfaces on both sides is measured here as approximately 1 nm, *i.e.* almost twice as much as the figures found by reflectivity. Convolution with the probe-size of around 0.1 nm cannot account for the entire discrepancy. EELS signals can suffer from delocalisation, *i.e.* excitation of an atom by a fast electron passing at a distance which can be several angstroms. However the effects would not be expected to be very large here, especially in the case of the high energy-loss Mg K edge. The broadening of the profiles is more likely to be due to non-negligible multiple elastic and inelastic scattering in the region where the data were acquired (~40 nm) which is still not ideally thin for 100 keV electrons and these materials, and the disordered layers seen to be present at the sample surface which almost certainly contain a mixture of Mg and Co. Further EELS experiments on thinner samples (or using higher acceleration voltages) and using higher energy resolution should provide further details of the structure and chemistry of the range of samples studied here.

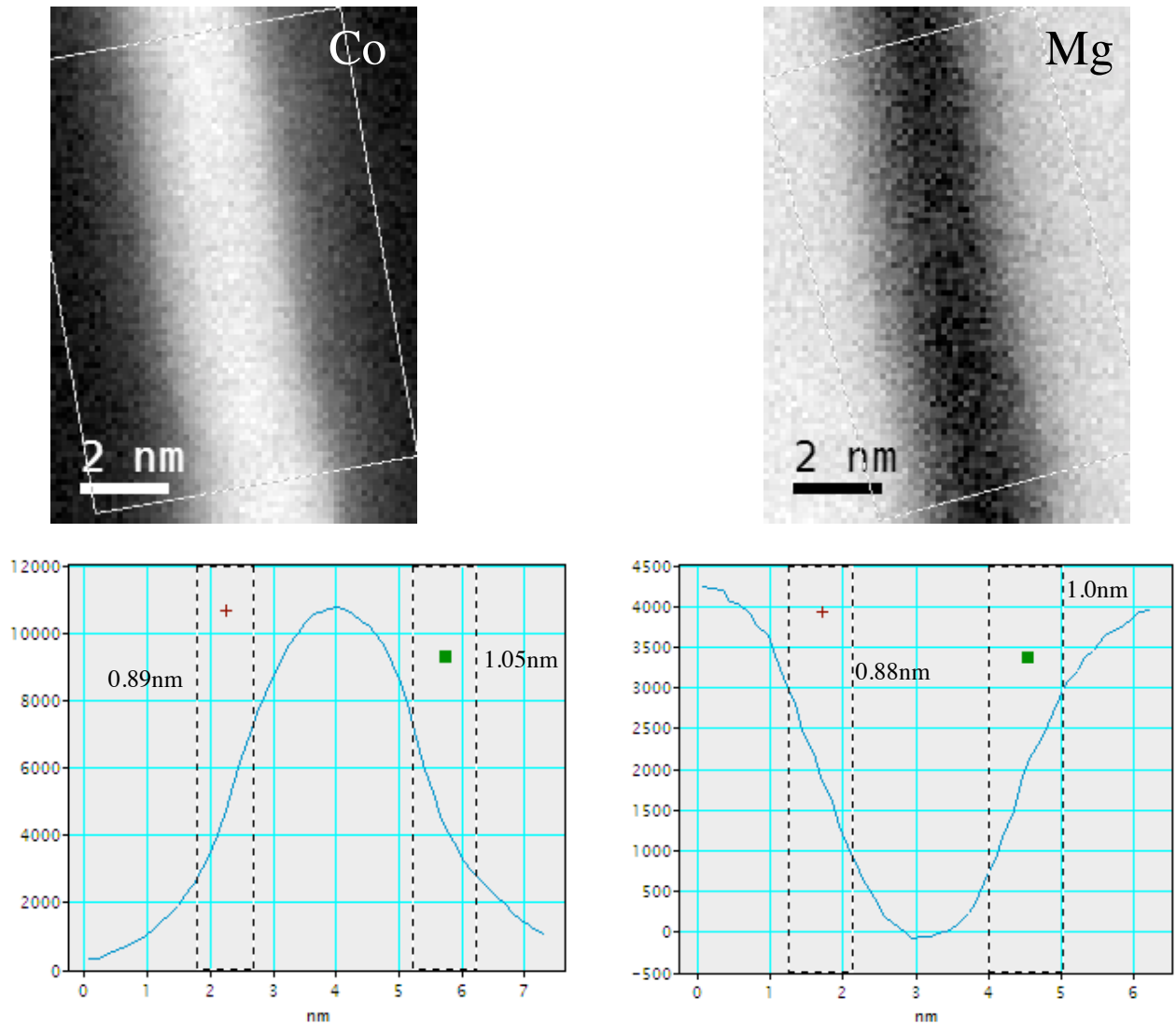

**Figure 4.** Higher resolution EELS maps (top) of Co (left) and Mg (right) across the second Co layer and the interface width profiles (bottom).

*3.2 Samples with annealing*

<u>*EUV reflectivity*</u>

We show in Figure 5 the reflectivity curves as a function of the annealing temperature, measured in the 45-55 eV energy range at 45° Bragg angle, angle for which the multilayer is designed, and in the 42-52 eV energy range at 50°. At 45° the maximum reflectivity is 42.4% at 49.3 eV, which is close to the Mg-L absorption edge at 49.50 eV for $2p_{3/2}$ and 49.78 eV for $2p_{1/2}$. This is rather too close to the Mg-L edge and we prefer to fit the reflectivity curves obtained at 50°. In the following, the position and maximum position of the Bragg peak will then be discussed for spectra recorded at 50° of grazing incidence.

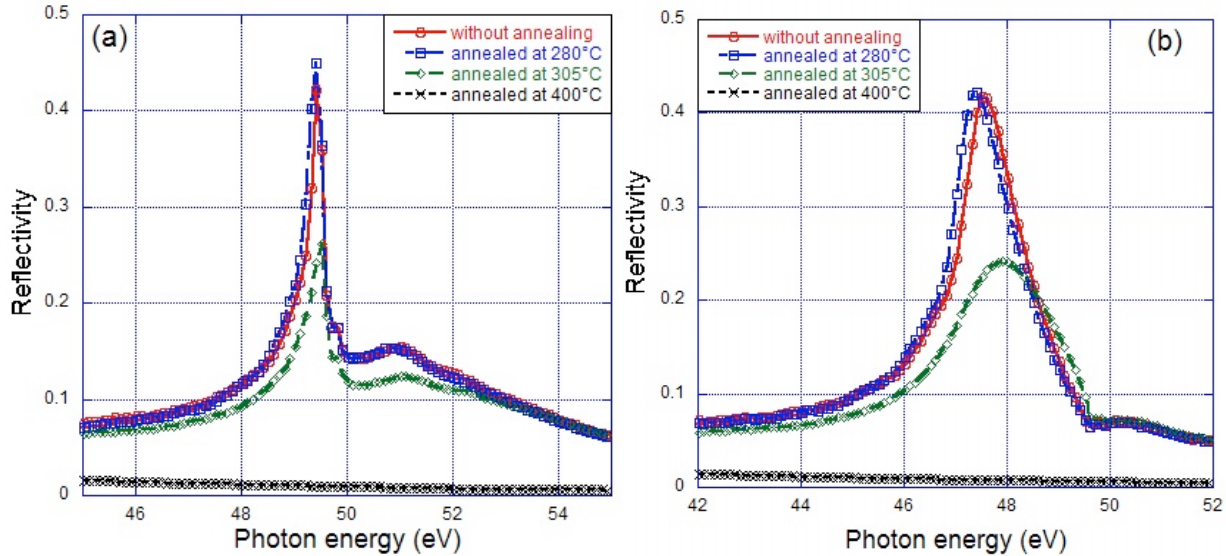

**Figure 5.** Evolution as a function of the annealing temperature of the Mg/Co_2 EUV reflectivity curves obtained at 45° (a) and 50° (b) Bragg angles.

The reflectivity slightly increases and the position of the Bragg peak of the Mg/Co_2 annealed at 280°C is slightly shifted (0.2 eV) towards lower photon energy with respect to that of the multilayer without annealing. At 305°C, a reflectivity drop of nearly 45% is observed as well as a peak shift (0.4 eV) towards higher photon energy. At 400°C, the reflectivity can be considered as zero. The structural parameters of the stack deduced from the fit of the reflectivity curves measured at 50° are collected in Table 2. The interfacial roughness and the thickness are stable for the unannealed and 280°C samples. However, the interfacial roughness increases quite a lot for the Mg/Co_2 annealed at 305°C and the fitted thicknesses vary drastically.

**Table 2.** Parameter values extracted from the fit to the Mg/Co_2 reflectivity curve measured at a Bragg angle of 50° in the EUV range. σ stands for the interfacial roughness.

| Multilayer Mg/Co_2 | d (nm) | $d_{co}$ (nm) $d_{Mg}$ (nm) | σ (nm) |
|---|---|---|---|
| Without annealing | 16.9 | 2.6<br>14.3 | Co-on-Mg = 0.5<br>Mg-on-Co = 0.6 |
| Annealed at 280°C | 16.98 | 2.86<br>14.12 | Co-on-Mg = 0.5<br>Mg-on-Co = 0.6 |
| Annealed at 305°C | 16.9 | 3.95<br>12.95 | Co-on-Mg = 1.0<br>Mg-on-Co = 2.7 |

### *NMR analysis*

NMR spectra obtained for the Mg/Co_1 multilayers as a function of the annealing temperatures are presented in Figure 7(a). As can be seen, the spectra shapes of the Mg/Co_1 multilayers after annealing at different temperatures show a well defined line at 226 MHz which corresponds to Co atoms lying in the bulk of the Co layers with an hcp structure. The Co atoms are thus mainly situated in pure Co layers (226 MHz) and the intermixing at the Mg/Co interfaces is limited. An additional line is observed at 156 MHz for the multilayer without annealing and for the sample annealed at 280°C in Figure 7(b). This line has been identified in our previous work with the Co atoms situated at perfect interfaces between Co and Mg atoms [1]. The overall shape of the spectrum after annealing at 280°C is very similar to that observed for the as deposited sample. It can be noticed that the Co interface line seems to be even better resolved after annealing at 280°C than in the spectrum of the as deposited sample. This observation would suggest that such low temperature annealing results in some sharpening of the Co/Mg interfaces.

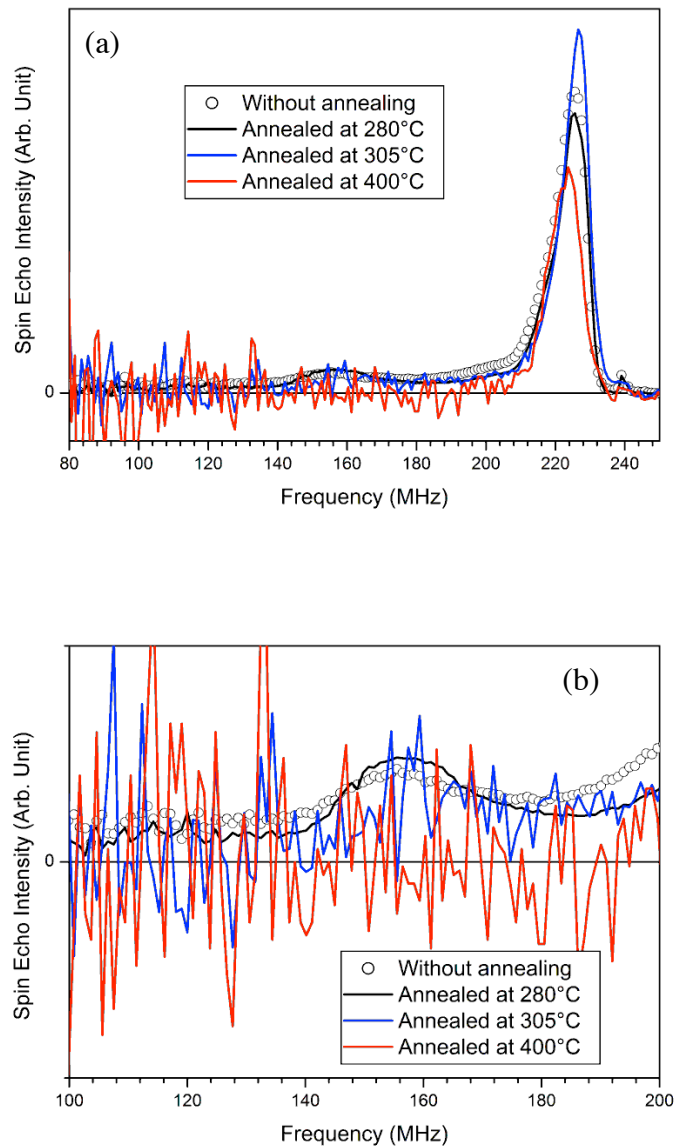

**Figure 6.** (a) NMR spectra of Mg/Co_1 multilayers as a function of the annealing temperatures; (b) zoom around 156 MHz.

The modification of the NMR spectra is much more significant for higher annealing temperatures (305 and 400°C). The most striking observation is that the contribution of the Co atoms at the Co/Mg interfaces (156 MHz) vanishes with increasing annealing temperature, Figure 7(b). This could be attributed to the formation of non magnetic interfacial alloys or compounds. However this is not consistent with the XES analysis and the observation that the total NMR intensity is not significantly reduced after annealing.

Additional details can be obtained from the magnetic information that is provided by the NMR experiments. Indeed NMR can also probe the magnetic stiffness of the samples [16]. This stiffness can have several origins and is described as a magnetic restoring field. The plot of this restoring field versus the annealing temperature is shown in Figure 8 (room temperature corresponding to 20°C). It can be seen that while the restoring field stays unchanged after the annealing at 280°C, it increases strongly for higher annealing temperatures. After the annealing at 400°C the restoring field is an order of magnitude higher than after the annealing at 280°C. This observation does not support the intermixing process since such a process usually leads to a decrease in the restoring field. However it shows that the magnetization processes are drastically modified by the annealing process.

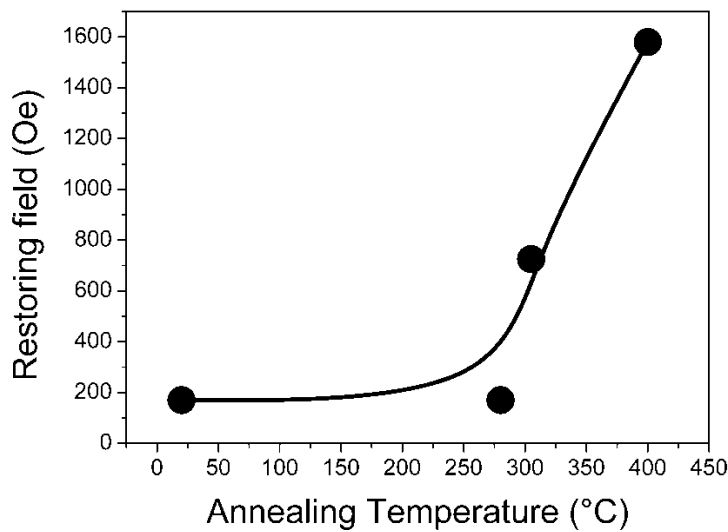

**Figure 7.** Magnetic restoring field versus the annealing temperature of Mg/Co_1 multilayers.

Actually the vanishing of the interfacial line as well as the increase in the restoring field with the annealing temperature can both be explained by the clustering of the Co layers into spheres or more probably into Co grains with pan cake like shapes. Indeed spherical grains would lead to a reorientation of the magnetization direction of the Co grains compared to the magnetization direction of the Co thin film. In such a case the NMR resonance frequency would shift from 226 MHz to 220 MHz [31], which is not observed in the NMR spectra. The clustering of the Co atoms will result in Co particles for which the surface to volume ratio is much smaller than in the case for the as deposited multilayer shape. If the Co grains are large enough, the contribution of the surface Co atoms becomes so small that it is not measurable anymore. The magnetic properties of Co grains are also very different from those of Co films. In Co films extended domain wall (Neel walls) can develop, resulting in soft magnetic properties (small restoring fields in NMR) while in Co grains such domain walls if they exist will have a smaller mobility, resulting in hard magnetic properties (large restoring fields in NMR). Therefore the clustering of the Co films into large Co grains with increasing annealing temperature explains both the evolution of the NMR spectra and the evolution of the magnetic properties of the samples.

*XES analysis*

We present in Figure 8 the Mg Kß emission band of the Mg/Co_1 sample as a function of the annealing temperature and the comparison to the references Mg and MgO. The Mg Kß emission band of the Mg/Co multilayer without annealing is discussed in detail in Ref. [1].

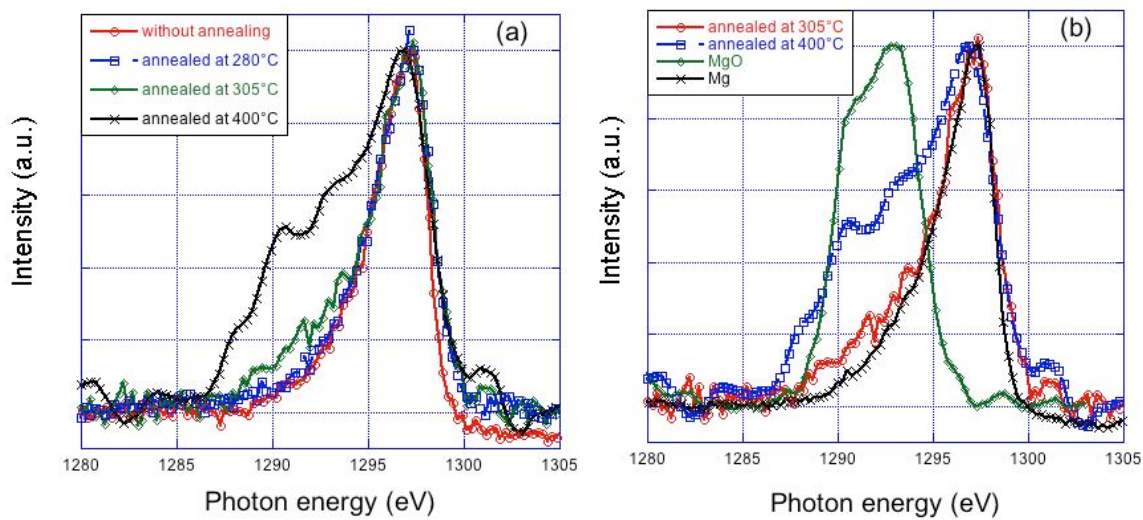

**Figure 8.** The Mg Kß emission band of sample Mg/Co_1 annealed at different temperatures (a) and its comparison with the Mg and MgO spectra (b).

The spectrum of the sample annealed at 280°C is close that obtained without annealing and close to that of Mg metal. We deduce that within the sample annealed at 280°C, the Mg atoms are in a physico-chemical state close to that of Mg atoms in the Mg metallic state and the interaction with the Co layers is limited. The spectra of Mg/Co_1 samples annealed at 305°C and 400°C differ from that of Mg/Co without annealing toward the low photon energies. A shoulder appears in the region of the MgO maximum (1288-1295 eV) whose intensity increases with temperature. Thus from 305°C an oxidation of the Mg layer takes place. Because of the annealing conditions, the oxidation takes place after the EUV measurements when the sample is returned to the air prior their XES analysis. Thus, whatever the annealing temperature, the Mg atoms do not interact with the Co atoms, which is in agreement with the NMR result.

### *SEM analysis*

From the SEM images of cross sections presented in Figure 9 we can see that the Mg/Co_2 multilayers without annealing and annealed at 280°C have clear multilayer structures. The cross section of the samples presents from top to bottom the Pt protection layer, the Mg/Co_2 multilayer and the Si substrate. For the sample annealed at 305°C, there exist some cracks between the multilayer and the substrate, the interface is no longer flat and there seems to be a loss of adherence. The sample annealed at 400°C has more serious changes compared to the sample annealed at 305°C: cracks and voids exist also between the layers. The change in the multilayer structure results from the change in the morphology of the Co layers deduced from NMR. The deformation of the Co layers generates mechanical stress that finally is discharged by the delamination of the multilayer and the formation of cracks. Because of these cracks, the multilayer is no longer flat and so its reflectivity decreases. Finally, the cracks facilitate the introduction of oxygen within the multilayer. This explains the oxidation of the Mg layers observed in XES.

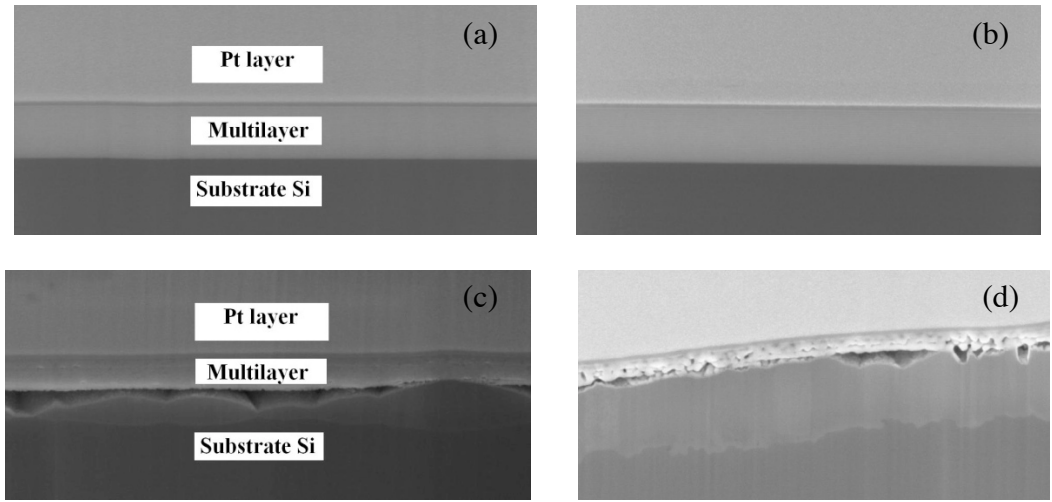

**Figure 9.** SEM images of Mg/Co_2 multilayer without annealing (a), annealed at 280°C (b), 305°C (c) and 400°C (d).

## 4. Conclusions

We have characterized a Mg/Co periodic multilayer and its thermal stability by using STEM / EELS, EUV reflectivity around 25 nm (~50 eV), NMR, XES and SEM. The multilayer is designed for optical application in the EUV range. STEM / EELS images show the good quality of the as-prepared multilayer and the crystallization of the Co and Mg layers. EUV reflectivity measurements show that the reflectivity slightly increases with the temperature up to 280°C. Then, the reflectivity drops by nearly 45% at 305°C and vanishes at 400°C. The combination of the results obtained by the various used techniques allows us to explain this reflectivity variation: as a function of the annealing temperature the morphology of the Co layers changes; this generates some cracks within the multilayer and then a loss of its flatness; consequently the reflectivity decreases until the multilayer is destroyed.

ACKNOWLEDGMENT

The authors from Tongji University are indebted to the National Natural Science Foundation of China (Granted No. 10825521 and 11061130549). Part of this work was made within the framework of the

REFERENCES

[1] K. Le Guen, M.-H. Hu, J.-M. André, P. Jonnard, S. K. Zhou, H. Ch. Li, J. T. Zhu, Z. S. Wang, C. Meny: J. Phys. Chem. C, 2010, 114, 6484.

[2] K. Le Guen, M.-H. Hu, J.-M. André, P. Jonnard, S. K. Zhou, H. Ch. Li, J. T. Zhu, Z. S. Wang, N. Mahne, A. Giglia, S. Nannarone: Applied Physics A, 2011, 102, 69.

[3] J. Zhu, S. Zhou, H. Li, Q. Huang, Z. Wang, K. Le Guen, M.-H. Hu, J.-M. André, P. Jonnard: Appl. Opt., 2010, 49, 3922.

[4] H. Maury, P. Jonnard, K. Le Guen, J. M. André, J. Wang, J. Zhu, J. Dong, Z. Zhang, F. Bridou, F. Delmotte, C. Hecquet, N. Mahne, A. Giglia, and S. Nannarone: European Physical Journal B, 2008, 64, 193.

[5] H. Takenaka, S. Ichimaru, T. Ohchi, E.M. Gullikson: J. Elec. Spec. Rel. Phenom., 2005, 144, 1047.

[6] T. Djavanbakht, V. Carrier, J.-M. Andre, R. Barchewitz and P. Troussel: J. Phys. IV France, 2000, 10, 10-281.

[7] P. Mengucci, G. Majni, A. Di Cristoforo, R. Checchetto, A. Miotello, C. Tosello and G. Principi: Thin Solid Films, 2003, 433, 205.

[8] H. -J. Stock, U. Kleineberg, B. Heidemann, K. Hilgers, A. Kloidt, B. Schmiedeskamp, U. Heinzmann, M. Krumrey, P. Müller and F. Scholze: Appl. Phys A, 1994, 58, 371.

[9] U. Kleineberg, H.-J. Stock, A. Kloidt, B. Schmiedeskamp, U. Heinzmann, S. Hopfe and R. Scholz:

Published in Applied Physics A 106, 737-745 (2012)physica status solidi (a), 1994, 145, 539.

[10] B. Heidemann, T. Tappe, B. Schmiedeskamp and U. Heinzmann: Z. Phys. B, 1995, 99, 37.

[11] R. Senderak, M. Jergel, S. Luby, E. Majkova, V. Holy, G. Haindl, F. Hamelmann, U. Kleineberg and U. Heinzmann: J. Appl. Phys., 1997, 81, 2229.

[12] M. Barthelmess, S. Bajt: Appl. Opt. 2011 50, 1610.

[13] M.-H. Hu, K. Le Guen, J.-M. André, P. Jonnard, S. K. Zhou, H. Ch. Li, J. T. Zhu, Z. S. Wang: AIP Conf. Proc., 2010, 1221, 56.

[14] S. Nannarone, F. Borgatti, A. DeLuisa, B. P. Doyle, G. C. Gazzadi, A. Giglia, P. Finetti, N. Mahne, L. Pasquali, M. Pedio, G. Selvaggi, G. Naletto, M. G. Pelizzo, G. Tondello: AIP Conference Proceedings, 2004, 708, 450.

[15] C. Meny, and P. Panissod: Modern Magnetic Resonance, G. Webb, Ed., Springer, Heidelberg, 2006.

[16] P. Panissod, and C. Meny: Applied Magnetic Resonance, 2000, 19, 447.

[17] C. Bonnelle, F. Vergand, P. Jonnard, J.-M. André, P. F. Staub, P. Avila, P. Chargelègue, M.-F. Fontaine, D. Laporte, P. Paquier, A. Ringuenet, B. Rodriguez: Review of Scientific Instruments, 1994, 65, 3466.

[18] H. Maury, P. Jonnard, J.-M. André, J. Gautier, M. Roulliay, F. Bridou, F. Delmotte, M.-F. Ravet, A. Jérome, P. Holliger: Thin Solid Films, 2006, 514, 278.

[19] M. Iwami, M. Kusaka, M. Hirai, R. Tagami, H. Nakamura, and H. Watabe: Applied Surface Science, 1997, 117, 434.

[20] E. Z. Kurmaev, V. R. Galakhov, and S. N. Shamin: Crit. Rev. Solid State Mater. Sci., 1998, 23,20

Published in Applied Physics A 106, 737-745 (2012)65.

[21] N. Miyata, S. Ishikawa, M. Yanagihara, and M. Watanabe: Japanese Journal of Applied Physics Part 1-Regular Papers Short Notes & Review Papers, 1999, 38, 6476.

[22] I. Jarrige, P. Jonnard, N. Frantz-Rodriguez, K. Danaie, and A. Bosseboeuf: Surface and Interface Analysis, 2002, 34, 694.

[23] V. R. Galakhov: X-Ray Spectrometry, 2002, 31, 203.

[24] M. Salou, S. Rioual, J. Ben Youssef, D. T. Dekadjevi, S. P. Pogossian, P. Jonnard, K. Le Guen, G. Gamblin, and B. Rouvellou: Surface and Interface Analysis, 2008, 40, 1318.

[25] K. Le Guen, G. Gamblin, P. Jonnard, M. Salou, J. Ben Youssef, S. Rioual, and B. Rouvellou: European Physical Journal-Applied Physics, 2009, 45.

[26] H. Maury, J. M. André, K. Le Guen, N. Mahne, A. Giglia, S. Nannarone, F. Bridou, F. Delmotte, and P. Jonnard: Surface Science, 2009, 603, 407.

[27] P. Jonnard, F. Vergand, C. Bonnelle, E. Orgaz, and M. Gupta: Physical Review B (Condensed Matter), 1998, 57, 12111.

[28] P. Jonnard, K. Le Guen, R. Gauvin, J.-F. Le Berre: Microsc Microanal., 2009, 15, 36.

[29] L.V. Azaroff: "X-ray Spectroscopy". McGraw-Hill Inc., 1974.

[30] C. Bonnelle: X-ray spectroscopy. Annu. Rep. Prog. Chem., Sect. C, Phys. Chem., 1987, 84, 201.

[31] M. Malinowska, C. Meny, E. Jedryka, P. J. Panissod: Phys.: Condens. Matter. 1998, 10, 4919.
21